\pdfoutput=1

\documentclass[prl,twocolumn,showpacs,preprintnumbers,amsmath,amssymb,
superscriptaddress,nofootinbib]{revtex4-1}
\usepackage{epsfig}
\usepackage{graphicx}

\newcommand{\smallminus}{{\rm\rule[2.4pt]{6pt}{0.65pt}}}
\newcommand{\smallplus}{\hspace{0.5pt}\text{{\small+}}\hspace{-0.5pt}}
\newcommand{\mi}{\smallminus}
\newcommand{\psl}{\smallplus}

\newcommand{\ab}[1]{\langle #1\rangle}
\renewcommand{\sb}[1]{[ #1]}
\newcommand{\eq}[1]{\vspace{-0.15cm}\begin{equation}#1\vspace{-0.15cm}\end{equation}}

\begin{document}

\preprint{CALT-TH-2014-154}

\title{On the Singularity Structure of Maximally Supersymmetric Scattering Amplitudes}
\author{Nima Arkani-Hamed}\affiliation{School of Natural Sciences, Institute for Advanced Study, Princeton, NJ}
\author{Jacob L. Bourjaily}\affiliation{Niels Bohr International Academy and Discovery Center, Copenhagen, Denmark}
\author{Freddy Cachazo}\affiliation{Perimeter Institute for Theoretical Physics, Waterloo, Ontario, Canada}
\author{Jaroslav Trnka}\affiliation{Walter Burke Institute for Theoretical Physics, California Institute of Technology, Pasadena, CA}
\date{\today}

\begin{abstract}

We present evidence that loop amplitudes in maximally supersymmetric (\mbox{$\mathcal{N}\!=\!4$}) Yang-Mills (SYM) beyond the planar limit share some of the remarkable structures of the planar theory. In particular, we show that through two loops, the four-particle amplitude in full \mbox{$\mathcal{N}\!=\!4$} SYM has only logarithmic singularities and is free of any poles at infinity---properties closely related to uniform transcendentality and the UV-finiteness of the theory. We also briefly comment on implications for maximal ($\mathcal{N}\!=\!8$) supergravity.\\[-12pt]
\end{abstract}

\maketitle

\section{Introduction}\vspace{-10pt}
Recent years have seen enormous advances in our understanding of the structure of scattering amplitudes (see e.g.\ \cite{Dixon:1996wi,Cachazo:2005ga,Drummond:2011ic,Elvang:2013cua,Henn:2014yza,Beisert:2010jr} for reviews). Most of this progress has been for the case of \mbox{$\mathcal{N}\!=\!4$} SYM in the planar limit. In this theory, a reformulation of perturbation theory exists in which tree- and loop-amplitudes are both directly represented in terms of on-shell diagrams which can be computed as contour integrals over the (positive) Grassmannian, $G_+(k,n)$, \cite{ArkaniHamed:2009dn,ArkaniHamed:2012nw}. Moreover, the entire S-matrix can be defined geometrically in terms of a space called the `amplituhedron', \cite{Arkani-Hamed:2013jha,Arkani-Hamed:2013kca}. Unlike Feynman diagrams, on-shell diagrams do not represent local processes in space-time; but they make the Yangian symmetry of the theory completely manifest. 

Should we expect such magic beyond the planar limit? One school of thought is that the remarkable structures seen in planar \mbox{$\mathcal{N}\!=\!4$} SYM are all a consequence of the integrability of the theory, a feature that is certainly lost beyond the planar limit. We find this view rather implausible. Indeed, while integrability has been marvelously exploited in the computation of Wilson loops and the integrated amplitudes \cite{Basso:2013vsa,Basso:2013aha}, it has played essentially no direct role in unveiling the structures alluded to above in the structure of loop integrands. In particular, the emergence of the Yangian symmetry is an incidental observation in this picture, not central to any of the Grassmannian geometry underlying the integrand. The geometry underlying the form of loop integrands directly encodes their structure of singularities, and it is quite plausible that an on-shell reformulation of loop amplitudes may exist for any theory. 

Even in the planar limit, we know that amplitudes satisfy relations between different color-ordered sectors---such as the $U(1)$ decoupling and KK identities, \cite{Kleiss:1988ne}, or the BCJ relations \cite{Bern:2008qj,Bern:2010ue,Bern:2010fy}---which are not obvious in the new formulations. This also suggests that, instead of losing magic beyond the planar limit, we could expect to find even richer structures that both unify different color orderings and explain their interrelations in the planar limit, and also control structure beyond the planar limit. 

Studying the singularities of scattering amplitudes in the planar limit has played an important role in the discovery of hidden structures, and it is natural to continue this exploration beyond the planar limit. An important feature of planar \mbox{$\mathcal{N}\!=\!4$} SYM, which can be stated without referring to the color ordering or cyclic symmetry, is that the amplitudes have only logarithmic singularities, with no poles at infinity. This would follow from the existence of an on-shell reformulation of \mbox{$\mathcal{N}\!=\!4$} SYM beyond the planar limit, if such a formulation were found to exist. And so it is natural to conjecture that this property continues to hold for the full theory:
\vspace{-0pt}
\begin{center}\boxed{\parbox{0.95\columnwidth}{To all orders of perturbation theory, scattering amplitudes in \mbox{$\mathcal{N}\!=\!4$} SYM beyond the planar limit have only logarithmic singularities, without any poles at infinity.}}\end{center}\vspace{-0pt}

In this note we present evidence for this highly non-trivial property by showing that it holds for the two-loop four-particle amplitude in full \mbox{$\mathcal{N}\!=\!4$} SYM. \\[-20pt]

\section{Logarithmic Differential Forms}\vspace{-10pt}
Loop amplitudes in quantum field theory can be represented as integrals over some rational form on the space of loop momenta. We say that a rational form $\Omega$ is {\it logarithmic} if it has only logarithmic singularities; that is, if near any pole parameterized by $\epsilon\!\to\!0$, there exists a change of variables for which $\Omega$ locally takes the form $\widetilde{\Omega}\wedge d\epsilon/\epsilon$, where $\widetilde{\Omega}$ is a co-dimension one form with only logarithmic singularities independent of $\epsilon$. For any such differential form $\Omega$, there exists a change of variables $\{\ell_1,\ldots,\ell_L\}\!\mapsto\!\{\alpha_1,\ldots,\alpha_{4L}\}$, for which $\Omega$ takes the form,\\[-10pt]
\eq{\Omega=\frac{d\alpha_1}{\alpha_1}\wedge\cdots\wedge\frac{d\alpha_{4L}}{\alpha_{4L}}=d\!\log(\alpha_1)\wedge\cdots\wedge d\!\log(\alpha_{4L})}
(or a linear combination of differential forms of this type). We will refer to measures of this type as ``$d\!\log$-forms''.

We can see examples of such differential forms in the case of one-loop amplitudes, which can be decomposed into scalar bubble, triangle, and box integrals (see e.g.\ \cite{tHooft:1978xw}), corresponding to the integration measures:\\[-14pt]
\begin{align}&\hspace{-40pt}\mathcal{I}_2(\ell)\equiv\frac{d^4\ell}{\ell^2(\ell+p_2+p_3)^2};\quad\mathcal{I}_3(\ell)\equiv\frac{d^4\ell\,(p_1+p_2)^2}{\ell^2(\ell+p_2)^2(\ell-p_1)^2};\hspace{-40pt}\nonumber\\&\hspace{-40pt}\mathcal{I}_4(\ell)\equiv\frac{d^4\ell\,(p_1+p_2)^2(p_2+p_3)^2}{\ell^2(\ell+p_2)^2(\ell+p_2+p_3)^2(\ell-p_1)^2}.\end{align}
While the bubble integration measure is {\it not} logarithmic, it is known (see e.g.\ \cite{ArkaniHamed:2012nw}) that the box can be written in $d\!\log$-form, $\mathcal{I}_4(\alpha)\!=\!d\!\log(\alpha_1)\wedge\cdots\wedge d\!\log(\alpha_4)$, via:\\[-8pt]
\eq{\hspace{-70pt}\begin{array}{ll}\alpha_1\!\equiv\!\ell^2/(\ell\,\mi\,\ell^*)^2,&\alpha_3\!\equiv\!(\ell\psl\,p_2\psl\,p_3)^2/(\ell\,\mi\,\ell^*)^2,\\\alpha_2\!\equiv\!(\ell\psl\,p_2)^2/(\ell\,\mi\,\ell^*)^2,&\alpha_4\!\equiv\!(\ell\,\mi\,p_1)^2/(\ell\,\mi\,\ell^*)^2,\end{array}\hspace{-50pt}}
where $\ell^*\!\equiv\frac{\ab{23}}{\ab{31}}\lambda_1\widetilde\lambda_2$ is one of the quad-cuts of the box. Similarly, the triangle can also be written in $d\!\log$-form, $\mathcal{I}_3(\alpha)\!=\!d\!\log(\alpha_1)\wedge\cdots\wedge d\!\log(\alpha_4)$, via:
\eq{\hspace{-65pt}\alpha_1\!\equiv\!\ell^2,\;\alpha_2\!\equiv\!(\ell\psl\,p_2)^2,\;\alpha_3\!\equiv\!(\ell\,\mi\,p_1)^2,\;\alpha_4\!\equiv\!(\ell\cdot\ell^*),\hspace{-50pt}}
where $\ell^*\!\equiv\!\lambda_1\widetilde\lambda_2$.

Notice that while both the triangle and box integrals are logarithmic, only the box is free of a pole at \mbox{$\ell\!\mapsto\!\infty$}. And while both integrals are UV-finite (unlike the bubble), poles at infinity could possibly signal bad UV behavior. Although the absence of poles at infinity may not be strictly necessary for finiteness, 
the amplitudes for both $\mathcal{N}=4$ SYM and \mbox{$\mathcal{N}\!=\!8$} SUGRA are remarkably free of such poles through at least two-loops.

There are many reasons to expect that loop amplitudes which are logarithmic have uniform (maximal) transcendentality; and integrands free of any poles at infinity are almost certainly UV-finite. This makes it natural to to ask whether these properties can be seen term-by-term at the level of the integrand.

\section{Logarithmic Form of the Two-Loop Four-Point Amplitude in \mbox{$\mathcal{N}\!=\!4$} and \mbox{$\mathcal{N}\!=\!8$}}\vspace{-10pt}

Our experience with planar $\mathcal{N}=4$ SYM suggests that the natural representation of the integrand which makes logarithmic singularities manifest in terms of on-shell diagrams, which are not in general manifestly local term-by-term. However at low loop-order, it has also been possible to see logarithmic singularities explicitly in 
particularly nice local expansions \cite{ArkaniHamed:2010gh,Bourjaily:2013mma}. Since we don't yet have an on-shell reformulation of `the' integrand beyond the planar limit (which may or may not be clearly defined for non-planar amplitudes) we will content ourselves here with an investigation of the singularity structure starting with known local expansions of two-loop amplitudes. 

The four-point, two-loop amplitude in \mbox{$\mathcal{N}\!=\!4$} SYM and \mbox{$\mathcal{N}\!=\!8$} SUGRA has been known for some time, \cite{Bern:1997nh}. It is usually given in terms of two integrand topologies---one planar, one non-planar---and can be written as follows:
\eq{\hspace{-62.5pt}\mathcal{A}_{4,\mathcal{N}}^{2\text{-loop}}\!\!=\frac{\mathcal{K}_{\mathcal{N}}}{4}\!\!\!\sum_{\sigma\in\mathfrak{S}_4}\!\!\!\int\!\!\Big[\!C_{\sigma,\mathcal{N}}^{(P)}\mathcal{I}_\sigma^{(P)}\!\psl C_{\sigma,\mathcal{N}}^{(NP)}\mathcal{I}_\sigma^{(NP)}\!\Big]\hspace{-1pt}\delta^{4|2\mathcal{N}}\!\big(\lambda\!\cdot\!q\big)\hspace{-50pt}\vspace{-0.1cm}\label{two_loop_amplitude}}
where $\sigma$ is a permutation of the external legs and \mbox{$\delta^{4|2\mathcal{N}}(\lambda\!\cdot\!q)$} encodes super-momentum conservation with $q\!\equiv\!(\widetilde\lambda,\widetilde\eta)$; the factors $\mathcal{K}_{\mathcal{N}}$ are the permutation-invariants,\\[-10pt] 
\eq{\mathcal{K}_{4}\equiv\frac{\sb{3\,4}\sb{4\,1}}{\ab{1\,2}\ab{2\,3}}\quad\mathrm{and}\quad\mathcal{K}_8\equiv\left(\frac{\sb{3\,4}\sb{4\,1}}{\ab{1\,2}\ab{2\,3}}\right)^2\!\!;}
the integration measures $\mathcal{I}_\sigma^{(P)}\!,\,\mathcal{I}_\sigma^{(NP)}$ correspond to,\\[-10pt]
\vspace{-0.2cm}\begin{align}&\hspace{-15pt}\mathcal{I}_{1,2,3,4}^{(P)}\equiv (p_1+p_2)^2\times\hspace{-20pt} \raisebox{-40.5pt}{\includegraphics[scale=1]{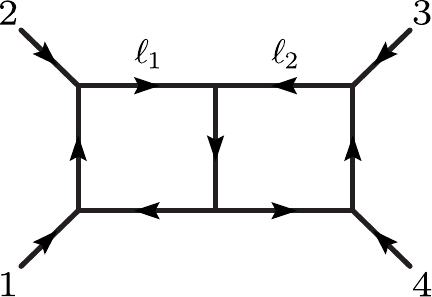}}\\[-7pt]&\hspace{-30pt}\mathrm{and}\nonumber\\[-7pt]&\hspace{-15pt}\mathcal{I}_{1,2,3,4}^{(NP)}\equiv (p_1+p_2)^2\times\hspace{-20pt} \raisebox{-40.5pt}{\includegraphics[scale=1]{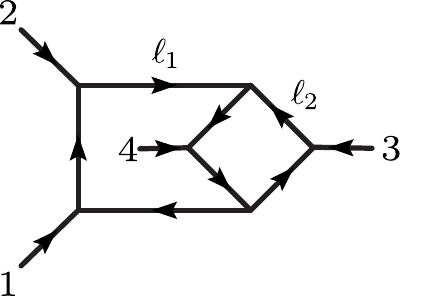}}\\[-20pt]\nonumber\end{align}
for $\sigma\!=\!\{1,2,3,4\}$; and the coefficients $C_{\{1,2,3,4\},\mathcal{N}}^{(P),(NP)}$ are the color-factors constructed out of structure constants $f^{abc}$'s according to the diagrams above for \mbox{$\mathcal{N}\!=\!4$}, and are both equal to $(p_1+p_2)^2$ for \mbox{$\mathcal{N}\!=\!8$}.

While the representation (\ref{two_loop_amplitude}) is correct, it obscures the fact that the amplitudes are ultimately logarithmic, maximally transcendental, and free of any poles at infinity. This is because the non-planar integral's measure, $\mathcal{I}_\sigma^{(NP)}$, is not itself logarithmic. We will show this explicitly below by successively taking residues until a double-pole is encountered; but it is also evidenced by the fact that its evaluation (using e.g.\ dimensional regularization) is not of uniform transcendentality, \cite{Tausk:1999vh}. These unpleasantries are of course cancelled in combination, but we would like to find an alternate representation of (\ref{two_loop_amplitude}) which makes this fact manifest term-by-term. Before providing such a representation, let us first show that the planar double-box integrand can be put into $d\!\log$-form, and then describe how the non-planar integrands can be modified in a way which makes them manifestly logarithmic.\\[-20pt]

\subsection{The Planar Double-Box Integral $\mathcal{I}^{(P)}_{\sigma}$}\vspace{-10pt}
In order to write $\mathcal{I}_{1,2,3,4}^{(P)}$ in $d\!\log$-form, we should first normalize it to have unit leading singularities. This is accomplished by rescaling it according to: \mbox{$\widetilde{\mathcal{I}}^{(P)}_{1,2,3,4}\!\equiv\!s\,t\,\mathcal{I}_{1,2,3,4}^{(P)}$}, where \mbox{$s\!\equiv\!(p_1\psl\,p_2)^2$} and \mbox{$t\!\equiv\!(p_2\psl\,p_3)^2$} are the usual Mandelstam invariants. Now that it is properly normalized, we can introduce an ephemeral extra propagator by multiplying the integrand by \mbox{$(\ell_1\psl\,p_3)^2/(\ell_1\psl\,p_3)^2$}, and notice that $\widetilde{\mathcal{I}}^{(P)}_{1,2,3,4}$ becomes the product of two boxes---motivating the following change of variables $\{{\ell}_1,{\ell}_2\}\!\to\!\{\alpha_1,\ldots,\alpha_8\}$:\\[-8pt]
\eq{\hspace{-155pt}\begin{array}{l@{$\,$}l}
\alpha_1\!\equiv\!(\ell_1\mi\,p_1\mi\,p_2)^2/(\ell_1\mi\,\ell_1^*)^2,&\alpha_5\!\equiv\!(\ell_1\psl\,\ell_2)^2/(\ell_2\mi\,\ell_2^*)^2,\\
\alpha_2\!\equiv\!(\ell_1\,\mi\,p_2)^2/(\ell_1\mi\,\ell_1^*)^2,&\alpha_6\!\equiv\!\ell_2^2/(\ell_2\mi\,\ell_2^*)^2,\\
\alpha_3\!\equiv\!\ell_1^2/(\ell_1\mi\,\ell_1^*)^2,&\alpha_7\!\equiv\!(\ell_2\mi\,p_3)^2/(\ell_2\mi\,\ell_2^*)^2,\\
\alpha_4\!\equiv\!(\ell_1\psl\,p_3)^2/(\ell_1\mi\,\ell_1^*)^2,&\alpha_8\!\equiv\!(\ell_2\mi\,p_3\mi\,p_4)^2/(\ell_2\mi\,\ell_2^*)^2,\end{array}\hspace{-150pt}}
where 
\eq{\ell_1^*\equiv\frac{\ab{1\,2}}{\ab{1\,3}}\lambda_3\widetilde\lambda_2\quad\mathrm{and}\quad\ell_2^*\equiv p_3+\frac{\!\!\phantom{{}^2}(\ell_1\psl\,p_3)^2\!\!}{\langle 4|\ell_1|3]}\lambda_4\widetilde\lambda_3,}
for which $\widetilde{\mathcal{I}}_{1,2,3,4}^{(P)}$ becomes \mbox{$d\!\log(\alpha_1)\wedge\cdots\wedge d\!\log(\alpha_8)$}. Notice that this form also makes it clear that the planar double-box is free of any poles at infinity. 

The existence of a logarithmic form of $\widetilde{\mathcal{I}}^{(P)}_{\sigma}$ also makes it clear that the integral will have uniform (maximal) transcendentality. Indeed, $\widetilde{\mathcal{I}}^{(P)}$ was computed using dimensional regularization in \mbox{ref.\ \cite{Bern:1997nh}}, which makes this fact clear---providing further evidence that logarithmic integrals have uniform (maximal) transcendentality.\\[-22pt] 

\subsection{The Non-Planar Double-Box Integral $\mathcal{I}^{(NP)}_{\sigma}$}\vspace{-10pt}
To see that the non-planar integral's measure $\mathcal{I}^{(NP)}_{1,2,3,4}$ is not logarithmic, consider the following co-dimension 7 residue. First, take the co-dimension four residue cutting the box-part of the diagram involving $\ell_2$. This gives a Jacobian of $(\ell_1\hspace{-0.5pt}\cdot\hspace{-0.5pt}q)(\ell_1\hspace{-0.5pt}\cdot\hspace{-0.5pt}\bar{q})$, where $q\!\equiv\!\lambda_3\widetilde\lambda_4,\bar{q}\!\equiv\!\lambda_4\widetilde\lambda_3$, resulting in the four-form:
\eq{\frac{d^4\ell_1\,(p_1\psl\,p_2)^2}{\ell_1^2(\ell_1\mi\,p_2)^2(\ell_1\mi\,p_1\,\mi\,p_2)^2(\ell_1\!\cdot\!q)(\ell_1\!\cdot\!\bar{q})}.}
We can then take the co-dimension 3 residue obtained by first cutting the two propagators $\ell_1^2$ and $(\ell_1\mi\, p_2)^2$, and then cutting the subsequent Jacobian via $\ell_1\!\mapsto\! x\, p_2$, resulting in the following differential form over $x$:
\eq{\mathcal{I}_{1,2,3,4}^{(NP)}\to\frac{dx}{(x+1)\,x^2\,t\,u}.\label{explicit_double_pole}}

The existence of such a double pole ensures that no $d\!\log$-form for $\mathcal{I}^{(NP)}_{\sigma}$ exists, and this is reflected in the fact that it is not of uniform transcendentality---as shown (using dimensional regularization) in \mbox{ref.\ \cite{Tausk:1999vh}}.

To cure the double pole in $\mathcal{I}^{(NP)}_{1,2,3,4}$ seen in (\ref{explicit_double_pole}), we should introduce an $\ell_1$-dependent factor in the numerator. It turns out that the following will suffice,
\eq{\hspace{-70pt}\mathcal{I}^{(NP)}_{1,2,3,4}\!\mapsto\!\widetilde{\mathcal{I}}^{(NP)}_{1,2,3,4}\!\equiv\!-\frac{(\ell_1\psl\, p_3)^2\psl\,(\ell_1\psl\,p_4)^2}{(p_3\psl\,p_4)^2}\,\mathcal{I}^{(NP)}_{1,2,3,4},\hspace{-50pt}\label{new_non_planar_int}}
as the limit taken above will result in a numerator factor of $x$, curing measure of the double pole found above.

We will soon demonstrate that the modified integral $\widetilde{\mathcal{I}}^{(NP)}_{1,2,3,4}$ is logarithmic by writing it explicitly in $d\!\log$-form. But let us first show that replacing $\mathcal{I}^{(NP)}_{1,2,3,4}\!\!\mapsto\widetilde{\mathcal{I}}^{(NP)}_{1,2,3,4}$ in the formula for the amplitude, (\ref{two_loop_amplitude}), will not change the final result. To see this, observe that the modification (\ref{new_non_planar_int}) amounts to subtracting 
\eq{\hspace{-70pt}(p_3\psl\,p_4)^2\psl\,(\ell_1\psl\, p_3)^2\psl\,(\ell_1\psl\,p_4)^2\!\!=\ell_1^2\psl\,(\ell_1\mi\,p_1\mi\,p_2)^2\hspace{-50pt}}
from the numerator of $\mathcal{I}^{(NP)}_{1,2,3,4}$, which is equivalent to subtracting two new integrals---each of which removes one propagator from $\mathcal{I}^{(NP)}_{1,2,3,4}$. Labeling these additional integrals by the leg attached to a quadrivalent vertex---i.e.,\\[-6pt]
\vspace{-5pt}\eq{\mathcal{I}_6^{(1)}\equiv\raisebox{-32.5pt}{\includegraphics[scale=1]{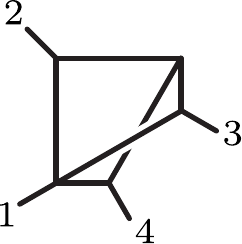}}\vspace{-0pt}}
we see that, 
\eq{\widetilde{\mathcal{I}}^{(NP)}_{1,2,3,4}=\mathcal{I}^{(NP)}_{1,2,3,4}-\mathcal{I}_6^{(1)}-\mathcal{I}_6^{(2)}.\label{two_loop_nonplanar_tilde}}
Thus, when each of the non-planar contributions of the amplitude are included according to equation (\ref{two_loop_amplitude}), the new integrand $\mathcal{I}_6^{(1)}$, for example, will contribute (twice) the following to the amplitude:\\[-10pt]
\eq{\mathcal{I}_6^{(1)}\Big(C_{\{1,2,3,4\},\mathcal{N}}^{(NP)}+C_{\{1,3,2,4\},\mathcal{N}}^{(NP)}+C_{\{1,4,2,3\},\mathcal{N}}^{(NP)}\Big).\label{correction_terms}} 
In \mbox{$\mathcal{N}\!=\!4$} SYM, these coefficients are color factors, and the combination appearing in (\ref{correction_terms}) exactly vanishes due to the Jacobi identity; for \mbox{$\mathcal{N}\!=\!8$}, the combination is simply $(s\psl\,t\psl\, u)$ which vanishes by momentum conservation. Thus, the replacing $\mathcal{I}_\sigma^{(NP)}$ with $\widetilde{\mathcal{I}}_\sigma^{(NP)}$ in the expression for the amplitude, (\ref{two_loop_amplitude}), does not change the result. 

Before moving on, we should point out that the correction integral $\mathcal{I}_6^{(a)}$ has been computed in \mbox{ref.\ \cite{Tausk:1999vh}} using dimensional regularization. From this, it is easy to verify that the combination appearing in (\ref{two_loop_nonplanar_tilde}) has uniform, maximal transcendentality. Of course, having uniform transcendentality is expected to follow from the existence of a $d\!\log$-form for $\widetilde{\mathcal{I}}^{(NP)}_{\sigma}$, but the explicit computation in \mbox{ref.\ \cite{Tausk:1999vh}} provides further evidence that this is indeed the case. \\[-22pt]

\subsection{Logarithmic Form of the Non-Planar Integral $\widetilde{\mathcal{I}}_{\sigma}^{(NP)}$}\vspace{-10pt}
Let us simply state the result. Unlike the planar double-box, $\widetilde{\mathcal{I}}_{1,2,3,4}^{(NP)}$ is a combination of $d\!\log$-forms:
\vspace{-3pt}\eq{\hspace{-65pt}\widetilde{\mathcal{I}}_{1,2,3,4}^{(NP)}=\Omega_1\wedge\left[\frac{1}{2}(c_1\psl\,c_2)\Omega_2^{(\mathrm{even})}\!\psl\frac{1}{2}(c_1\mi\,c_2)\Omega_2^{(\mathrm{odd})}\right].\hspace{-50pt}\vspace{-3pt}}
where $c_1\!\equiv\!1/(s\,t)$ and $c_2\!\equiv\!1/(s\,u)$, and $\Omega_1$ represents the $d\!\log$-form of the $\ell_2$ sub-integral, $d\!\log(\alpha_1)\wedge\cdots\wedge d\!\log(\alpha_4)$, with:
\vspace{-3pt}\eq{\hspace{-50pt}\begin{array}{l@{$\,$}l}\alpha_1\!\equiv\!(\ell_1\psl\,\ell_2)^2/(\ell_2\,\mi\,\ell_2^*)^2,&\alpha_3\!\equiv\!(\ell_2\mi\,p_3)^2/(\ell_2\,\mi\,\ell_2^*)^2,\\\alpha_2\!\equiv\!\ell_2^2/(\ell_2\,\mi\,\ell_2^*)^2,&\alpha_4\!\equiv\!(\ell_1\psl\,\ell_2\psl\,p_4)^2/(\ell_2\,\mi\,\ell_2^*)^2,\end{array}\hspace{-50pt}\nonumber\vspace{-3pt}}
where $\ell_2^*\!\equiv\!-\frac{\lambda_3\,(\ell_1|4\rangle)}{\ab{3\,4}}$ is the position of the quad-cut of the sub-box. As for the planar double-box, cutting the $\ell_2$ box introduces a Jacobian of $1/((\ell_1\hspace{-0.5pt}\cdot\hspace{-0.5pt}q)(\ell_1\hspace{-0.5pt}\cdot\hspace{-0.5pt}\bar{q}))$ with $q\!\equiv\!\lambda_3\widetilde\lambda_4$, $\bar{q}\!\equiv\!\lambda_4\widetilde\lambda_3$. This effectively makes the $\ell_1$ sub-integral a pentagon, with both parity-even and parity-odd contributions. The parity-even part can be considered the sum of three ``boxes'' according to:\\[-14pt]
\eq{\Omega_2^{(\mathrm{even})}\!\!\equiv\sum_{a=1}^3d\!\log(\beta_1^a)\wedge d\!\log(\beta_2^a)\wedge d\!\log(\beta_3^a)\wedge d\!\log(\beta_4^a)\nonumber\vspace{-8pt}}
where\\[-14pt]
\eq{\hspace{-50pt}\begin{array}{l@{$\;\;\;$}l}\beta_1^a\!\equiv\!(\ell_1\mi\,p_1\mi\,p_2)^2/(\ell_1\,\mi\,\ell_1^{*,a})^2,&\beta_3^a\!\equiv\!Q_a/(\ell_1\,\mi\,\ell_1^{*,a})^2,\\\beta_2^a\!\equiv\!\ell_1^2/(\ell_1\,\mi\,\ell_1^{*,a})^2,&\beta_4^a\!\equiv\!Q_{a+1}/(\ell_1\,\mi\,\ell_1^{*,a})^2,\end{array}\hspace{-50pt}\nonumber\vspace{-8pt}}
with $Q_a\!\equiv\!\{(\ell_1\mi\,p_2)^2,(\ell_1\!\cdot\!q),(\ell_1\!\cdot\!\bar{q})\}$ (with $Q_4\!\equiv\!Q_1$),\\[-10pt]
\eq{\ell_1^{*,1}\!\equiv\!\frac{\ab{3\,4}}{\ab{2\,3}}\lambda_2\widetilde\lambda_4,\quad\ell_1^{*,2}\!\equiv\!\mi\,p_4,\quad\ell_1^{*,3}\!\equiv\!\frac{\ab{1\,2}}{\ab{1\,4}}\lambda_4\widetilde{\lambda}_2.\vspace{-5pt}}
The parity-odd contribution is given by a single term, $\Omega_2^{(\mathrm{odd})}\!\equiv\!d\!\log(\gamma_1)\wedge\cdots\wedge d\!\log(\gamma_4)$, where:
\eq{\hspace{-50pt}\begin{array}{l@{$\;\;\;\;\;$}l}\gamma_1\!\equiv\!(\ell_1\mi\,p_1\mi\,p_2)^2/(\ell_1\mi\,p_2)^2,&\gamma_3\!\equiv\!(\ell_1\!\cdot\!q)/(\ell_1\mi\,p_2)^2,\\\gamma_2\!\equiv\!\ell_1^2/(\ell_1\mi\,p_2)^2,&\gamma_4\!\equiv\!(\ell_1\!\cdot\!\bar{q})/(\ell_1\mi\,p_2)^2.\end{array}\hspace{-50pt}\nonumber}

There are of course many equivalent ways of writing these $d\!\log$-forms. The fact that $\widetilde{\mathcal{I}}_\sigma^{(NP)}$ is not a {\it single} $d\!\log$-form reflects the fact that it does not have unit leading singularities---that is, that not all of its co-dimension eight residues are the same up to a sign. In fact, different contours give either $c_1$ or $c_2$ as their residue. This is interesting for \mbox{$\mathcal{N}\!=\!4$} SYM, because these become nothing but Parke-Taylor denominators for different orderings:
\vspace{-5pt}\eq{c_1\mathcal{K}_{4}=\frac{1}{\ab{1\,2}\ab{2\,3}\ab{3\,4}\ab{4\,1}},\; c_2\mathcal{K}_4=\frac{1}{\ab{1\,2}\ab{2\,4}\ab{4\,3}\ab{3\,1}}.\nonumber\vspace{-3pt}}

The fact that both of the integral's leading singularities are Parke-Taylor denominators (up to re-ordering) is more striking than it may at first appear: superconformal symmetry alone does not forbid leading singularities from having double-poles or even more complicated poles, involving differences of products of brackets, for example. And yet it can be shown, \cite{Nonplanar}, that for all multiplicity, all leading-singularities of MHV amplitudes to all loop-orders are simply combinations of differently-ordered Parke-Taylor denominators.\\[-22pt]

\vspace{-0pt}\section{Outlook and Conclusions}\vspace{-6pt}
We have seen sharp evidence that nonplanar amplitudes have the same kind of remarkable structure present in the planar limit, and that there are reasons to expect that this structure is closely related to uniform transcendentality and UV-finiteness. We have conjectured that this is possible to all orders of perturbation theory for \mbox{$\mathcal{N}\!=\!4$} SYM. 

Might the same result to hold for $\mathcal{N}=8$ SUGRA, as we have seen at two-loops? For \mbox{$\mathcal{N}\!=\!4$} SYM, the grassmannian structure of on-shell diagrams gives us forms with logarithmic singularities even beyond the planar limit \cite{ArkaniHamed:2012nw,Nonplanar}, which gives circumstantial support to the conjecture. This is not true for \mbox{$\mathcal{N}\!=\!8$} SUGRA leading singularities, and we don't have any particular reason to believe the statement to hold one way or another. Nonetheless the parallel with \mbox{$\mathcal{N}\!=\!4$} SYM at 2-loops is already a nice surprise, and motivates an exploration of  the singularity structure for  \mbox{$\mathcal{N}\!=\!8$} SUGRA integrands even at 3 and 4 loops where it is known that no divergences appear in the integrated amplitude, \cite{Bern:2007hh,Bern:2009kd}, in order to see whether this surprising parallel with the logarithmic singularity structure of \mbox{$\mathcal{N}\!=\!4$} SYM persists, potentially giving new insight into the question of the UV properties of the theory (see e.g.\ \cite{Bern:2009kd,Bjornsson:2010wm,Bern:2011qn}).

\acknowledgments We thank Zvi Bern, Lance Dixon, Enrico Herrmann, Sean Litsey, and James Stankowicz for helpful discussions. N.~A.-H. is supported by the Department of Energy under grant number DE-FG02-91ER40654; J.~B. is supported by a MOBILEX grant from the Danish Council for Independent Research; F.~C. is supported by the Perimeter Institute for Theoretical Physics which is supported by the Government of Canada through Industry Canada and by the Province of Ontario through the Ministry of Research \& Innovation; and J.~T. is supported in part by the David and Ellen Lee Postdoctoral Scholarship and by the Department of Energy under grant number DE-SC0011632.

\end{document}